\documentclass[conference]{IEEEtran}
\IEEEoverridecommandlockouts
\usepackage{amsmath,amssymb,amsfonts}
\usepackage{algorithmic}
\usepackage{graphicx}
\usepackage{textcomp}
\usepackage[table,xcdraw]{xcolor}
\usepackage{balance}
\usepackage{soul}
\usepackage{booktabs}
\usepackage{multirow}
\usepackage{float}
\usepackage{url}
\usepackage{tcolorbox}
\usepackage{tabularx} 
\usepackage[noadjust]{cite}
\usepackage[hidelinks,bookmarks=true,pagebackref=true]{hyperref}
\usepackage{subcaption}

\usepackage{enumitem}
\setlist[itemize,enumerate]{noitemsep, topsep=0pt, leftmargin=1.0em}
\usepackage[utf8]{inputenc}
\usepackage{newunicodechar,graphicx}
\DeclareRobustCommand{\okina}{%
  \raisebox{\dimexpr\fontcharht\font`A-\height}{%
    \scalebox{0.8}{`}%
  }%
}
\newunicodechar{ʻ}{\okina}

\newcommand{\RQA}{\textbf{RQ1}: At what frequency are assertion messages included in unit tests, and what is the rationale for including them?}
\newcommand{\RQAa}{\textbf{RQ1.1}:  How frequently do practitioners include an assertion message in their assertion methods?}
\newcommand{\RQAb}{\textbf{RQ1.2}:  What motivates practitioners to include/exclude assertion messages?}

\newcommand{\RQB}{\textbf{RQ2}: How do practitioners typically construct assertion messages in unit tests?}

\newcommand{\RQC}{\textbf{RQ3}: How are assertion messages maintained and kept relevant throughout the development lifecycle?}

\newcommand{\RQD}{\textbf{RQ4}: How are assertion messages integrated into the development and debugging processes?}

\begin{document}

\title{On the Rationale and Use of Assertion Messages in Test Code: Insights from Software Practitioners}

\author{\IEEEauthorblockN{Anthony Peruma}
\IEEEauthorblockA{\textit{University of Hawaiʻi at Mānoa} \\
Hawaiʻi, USA \\
peruma@hawaii.edu}
\and
\IEEEauthorblockN{Taryn Takebayashi}
\IEEEauthorblockA{\textit{University of Hawaiʻi at Mānoa} \\
Hawaiʻi, USA\\
tarynet@hawaii.edu}
\and
\IEEEauthorblockN{Rocky Huang}
\IEEEauthorblockA{\textit{University of Hawaiʻi at Mānoa} \\
Hawaiʻi, USA \\
rhuang8@hawaii.edu}
\and
\IEEEauthorblockN{Joseph Carmelo Averion}
\IEEEauthorblockA{\textit{University of Hawaiʻi at Mānoa} \\
Hawaiʻi, USA \\
averionj@hawaii.edu}
\and
\IEEEauthorblockN{Veronica Hodapp}
\IEEEauthorblockA{\textit{Columbia University} \\
New York, USA \\
vth2106@columbia.edu}
\and
\IEEEauthorblockN{Christian D. Newman}
\IEEEauthorblockA{\textit{Rochester Institute of Technology} \\
New York, USA \\
cnewman@se.rit.edu}
\and
\IEEEauthorblockN{Mohamed Wiem Mkaouer}
\IEEEauthorblockA{\textit{University of Michigan-Flint} \\
Michigan, USA \\
mmkaouer@umich.edu}
}

\maketitle

\begin{abstract}
Unit testing is an important practice that helps ensure the quality of a software system by validating its behavior through a series of test cases. Core to these test cases are assertion statements, which enable software practitioners to validate the correctness of the system's behavior. To aid with understanding and troubleshooting test case failures, practitioners can include a message (i.e., assertion message) within the assertion statement. While prior studies have examined the frequency and structure of assertion messages by mining software repositories, they do not determine their types or purposes or how practitioners perceive the need for or the usage of various types of assertion messages.

In this paper, we survey 138 professional software practitioners to gather insights into their experience and views regarding assertion messages. Our findings reveal that a majority of survey respondents find assertion messages valuable for troubleshooting failures, improving test understandability, and serving as documentation. However, not all respondents consistently include messages in their assertion methods. We also identified common considerations for constructing effective assertion messages, challenges in crafting them, maintenance techniques, and their integration into debugging processes.

Our results contribute to the understanding of current practices and provide guidelines for authoring high-quality assertion messages, serving as a foundation for best practices and coding standards. Furthermore, the insights can guide the improvement of automated unit testing tools by incorporating checks for the presence and quality of assertion messages and providing real-time feedback to practitioners.
\end{abstract}


\section{Introduction}

Unit testing is a vital part of software testing, where practitioners write code to validate the correctness of the software system through a series of test cases (i.e., test methods) \cite{pressman2005software}. Key to these test cases are assertion statements, also known as assertion (or assert) methods, that verify the expected behavior of the code under test. By comparing the actual and expected outcomes, the assertion methods determine whether a test case passes or fails \cite{hamill2004unit}. To help troubleshoot failed assertions and, to a certain extent, assist with comprehending the purpose of test cases, many unit testing frameworks provide practitioners with the ability to include an optional message parameter, known as an assertion message, which provides additional details about the failure \cite{meszaros2007xunit}.

The importance of assertion methods in ensuring the effectiveness of test suites has been well-established in prior work \cite{Lima2023ICSME,Zhang2015FSE}. Furthermore, multiple studies propose techniques to autogenerate assertion methods and show that these techniques result in assertions similar to developer-written methods \cite{Zamprogno2023,Midolo2023,Tufano2022,Watson2020}. However, assertion methods can also negatively impact code quality. Specifically, the test smell Assertion Roulette, where a test method contains multiple assertions without providing context for failures \cite{Bavota2012}, is associated with increased change-proneness in test code \cite{Spadini2018,Kim2020} and challenges in code comprehension \cite{Aljedaani2023}. These problems cannot be addressed without a stronger understanding of assertion messages, including how they are used and how they support comprehension. Therefore, the presence of assertion messages becomes critical in mitigating these side effects.

\subsection{Goal \& Research Questions}
\label{Section:Goal}
While studies that mine software repositories can empirically show the frequency and structure of these messages, they do not explain why practitioners structure their messages in specific ways or how practitioners interpret and perceive them in software quality and maintenance tasks \cite{Takebayashi2023NLBSE}. Therefore, the goal of this study is to \textit{gather insights directly from practitioners on the rationale, construction, maintenance, and effectiveness of assertion messages in software testing}. By understanding how practitioners approach assertion messages, including their motivations, practices, and challenges, we envision our findings leading to improvements in testing frameworks and the development of better tools and practices to support software testing and debugging, ultimately enhancing practitioner productivity and overall software quality.

We aim to answer the following research questions (RQs): 

\vspace{1.3mm}
\noindent\textbf{\RQA} While prior work exists on how frequently assertion messages appear in assertion methods \cite{Takebayashi2023NLBSE}, it is based on mining open-source software repositories and does not analyze the messages at the individual practitioners level, nor does it take into account the motivations and rationale behind practitioners' decisions to include or exclude assertion messages. As a human-subject study, this RQ, in comparison, reveals the extent to which individual practitioners prioritize and incorporate these messages into their test suites. Through this RQ, we gain insight into the perceived importance and value of these messages and the factors influencing practitioners' testing practices.

\vspace{1.3mm}
\noindent\textbf{\RQB} Through this RQ, we aim to better understand the practitioner's thought process for crafting an effective assertion message. The RQ explores the structure and semantics of the message and how they relate to the surrounding code to create informative and actionable messages. The findings will contribute to creating tools that automatically assess and recommend suitable assertion messages.

\vspace{1.3mm}
\noindent\textbf{\RQC} This RQ aims to understand the maintenance techniques practitioners employ to ensure that assertion messages are kept up-to-date and accurate. The findings can provide insights into unit testing best practices and overall improvements to software development and maintenance activities.

\vspace{1.3mm}
\noindent\textbf{\RQD} This RQ examines how practitioners utilize assertion messages to troubleshoot test case failures. The findings can lead to the development of tools and improvements in testing practices.  

\subsection{Contribution}
The main contributions of this work are:
\begin{itemize}
    \item Quantitative and qualitative insights from professional software engineers on their perception and experience of using assertion messages.
    \item Extending the body of knowledge on unit testing best practices and guidelines that improve the quality and maintainability of the test suite.
    \item Recommendations to guide the improvement of unit testing frameworks and code quality tools.
\end{itemize}

\section{Related Work}
\label{Section:related}
This section presents related studies on test code assertions and practitioner surveys. First, we describe studies examining assertion methods and messages. Next, we report on industry practitioner surveys on testing practices and challenges.

\subsection{Assertion Methods and Messages}
Zamprogno et al. \cite{Zamprogno2023} propose a dynamic, human-in-the-loop approach for generating assertion methods in test cases. Their tool, AutoAssert, focuses on generating assertions based on the runtime values of practitioner-selected variables. However, the generated assertions do not contain custom messages. The authors report that the generated assertions are similar to practitioner-written methods, and a user study shows that practitioners find the tool helpful. Tufano et al. \cite{Tufano2022} present an approach to generate accurate assertion statements using the pre-trained transformer model. Their model significantly outperforms prior techniques by pre-training on natural language and code and fine-tuning on a large dataset of assertion methods. Although the authors mention that their model can generate complex assertion methods, they do not discuss whether the generated assertion method contains a custom message. Watson et al. \cite{Watson2020} proposed a neural machine translation-based approach, called ATLAS, to generate meaningful assertion methods given test and focal methods. The authors evaluate ATLAS on a large corpus of open-source projects and report that ATLAS can predict the exact assertion method manually written by practitioners in 31\% of the cases. The authors do not discuss or mention anything about assertion messages in their study. Midolo and Tramontana \cite{Midolo2023} propose an approach to automatically generate test case templates that include JUnit assertions. The authors report that the generated test templates were clean and readable, making them easier for practitioners to understand and modify. They generated templates, increased code coverage, and helped detect faults. However, the authors do not discuss assertion messages in their study. Zhang and Mesbah \cite{Zhang2015FSE} conducted an empirical study that explored the relationship between test suite effectiveness and the number of assertions, assertion coverage, and different types of assertions. The authors report that assertion quantity and coverage are strongly correlated with effectiveness in detecting faults. 

An exploratory study by Takebayashi et al. \cite{Takebayashi2023NLBSE} analyzes assertion messages in 20 open-source Java projects to determine their characteristics and readability. The study reveals that practitioners rarely include messages in their assertions, with only about 5\% of assertions containing any messages. When messages are included, they are typically composed of string literals or a combination of string literals and identifiers. The authors use readability formulas to evaluate the messages and find that identifier-only messages require a beginner level of English to understand, whereas string literal messages require a 4th-grade reading level. In an analysis of 112 Java projects, Ma'ayan \cite{Maayan2018} reports that most projects do not adhere to unit testing best practices. The author reports that most projects violate the single-checkpoint principle, while only a small volume of projects adhere to the arrange, act, assert approach. The authors also report that most practitioners do not use custom assertion messages, making it difficult to understand the reason for test failures.  

\subsection{Practitioner Surveys}
Lima et al. \cite{Lima2023ICSME} survey 80 practitioners to understand their practices, benefits, and challenges regarding refactoring unit testing code. The authors highlight the reason for refactoring, which includes incorporating new assertion statements to verify newly added conditions. The authors also recommend that project teams emphasize the need to consider test code refactoring in planning activities and provide practitioners with the necessary resources to support these activities.
The practitioner survey by Kochhar et al. \cite{Kochhar2019} reports on factors considered important for test case quality across six dimensions. Among their findings, the authors highlight assertions' important role in identifying hidden defects. However, their study looks at assertions as a single category without examining the semantics of the messages. Runeson's \cite{Runeson2006} general survey of unit testing practices by practitioners highlights the strengths and weaknesses of unit testing. However, the author does not discuss assertion methods in the study. In their survey of 225 practitioners, Daka and Fraser \cite{Daka2014} provide insight into common practices and needs in unit testing. The authors show that there is a need to provide practitioners with best practices to write high-quality test cases. The authors also highlight the importance of providing practitioners with tools to help with test case generation and refactoring. Garousi and Zhi \cite{Garousi2013} conduct a survey among 246 software practitioners in Canada to identify trends, techniques, tools, and challenges in software testing. According to the authors, at the time of the survey, practitioners devote a significant amount of effort to unit testing, and traditional test-last development is more popular than test-driven development. The authors also note that maintaining automated test suites poses a challenge, as does refactoring overly rigid tests. Similar to other studies in this field, the authors emphasize the need for better training and education in software testing.

\noindent{\textbf{Summary.}} 
As described, prior studies have shown the importance of assertion methods in ensuring the effectiveness of test suites. However, only a few studies have examined the content and quality of assertion messages, from a data mining perspective, offering insights into their characteristics and readability. These studies do not examine how practitioners rely on these messages to resolve test case failures or how they use them in practice. As a result, there is a gap in understanding the practical value and implications of assertion messages, which is the focus of our study. 

\section{Study Design}
\label{Section:experiment_design}
In this section, we provide a comprehensive overview of our survey design, participant recruitment, and the methodology for analyzing responses. 
As our study involves working with human subjects, we made sure to address all ethical considerations. Prior to publishing the survey, we submitted it to the Institutional Review Board (IRB) of the Office of Research Compliance at our institution for review and approval.

\subsection{Survey Design}
We used Qualtrics \cite{Qualtrics} to create and host the survey and configured it to allow only one response per participant. Our survey consisted of 22 questions that were designed to gather information about the participants' demographics, technical experience, and experience with and perceptions of assertion messages. We determined these questions based on the overall goal of our study, discussed in Section \ref{Section:Goal}, and our review of related literature, discussed in Section \ref{Section:related}. Table \ref{Table:SurveyQuestions} shows the survey questions, the question type, whether a response is required, and any logic/notes applicable to each question. Additionally, before asking the unit testing questions, we provided simple examples of assertion methods in multiple programming languages to prevent/limit confusion or misunderstanding about an assertion method. We have included the complete questionnaire, as presented to participants, at \cite{Dataset}.

\begin{table*}[t]
\centering
\caption{Below are the questions that are part of the survey. The questionnaire, as presented to participants, that includes the answer options for the single-choice, multi-choice, and ranking questions, is available at \cite{Dataset}.}
\label{Table:SurveyQuestions}
\resizebox{\textwidth}{!}{%
\begin{tabular}{lp{0.56\linewidth}llp{0.2\linewidth}}
\toprule
\multicolumn{1}{c}{\textbf{No.}} &
  \multicolumn{1}{c}{\textbf{Question}} &
  \multicolumn{1}{c}{\textbf{Type}} &
  \multicolumn{1}{c}{\textbf{Required}} &
  \multicolumn{1}{c}{\textbf{Notes}} \\ \midrule
1 &
  What is your age range? &
  Single-Choice &
  Yes &
   \\ \midrule
2 &
  In which country do you currently reside? &
  Dropdown &
  Yes &
  Imported from Qualtrics question library \\ \midrule
3 &
  What is your primary language preference for both spoken and written communication? &
  Single-Choice &
  Yes &
  Includes ``Other'' free-text option \\ \midrule
4 &
  How do you describe yourself? &
  Single-Choice &
  Yes &
  Includes ``Self-describe'' free-text option \\ \midrule
5 &
  What best describes your current employment status? &
  Single-Choice &
  Yes &
  Includes ``Other'' free-text option \\ \midrule
6 &
  If you are currently employed/working, which of the following best describes your role in the organization? &
  Single-Choice &
  Yes &
  Includes ``Other'' free-text option \\ \midrule
7 &
  How many years of general programming experience do you have? &
  Single-Choice &
  Yes &
   \\ \midrule
8 &
  To what extent are you familiar with unit testing? &
  Single-Choice &
  Yes &
  End survey for ``Not familiar at all'' response \\ \midrule
9 &
  Which unit testing framework are you most experienced with? &
  Single-Choice &
  Yes &
  Includes ``Other'' free-text option \\ \midrule
10 &
  To what extent is writing unit tests part of your job/employment duties? &
  Single-Choice &
  Yes &
   \\ \midrule
11 &
  What is your primary Integrated Development Environment (IDE)? &
  Single-Choice &
  Yes &
  Includes ``Other'' free-text option \\ \midrule
12 &
  How frequently do you include an assertion error message in your assert method? &
  Single-Choice &
  Yes &
   \\ \midrule
13 &
  Please let us know why you selected this answer option. &
  Free Text &
  Yes &
  Shown only if ``Rarely'' or ``Never'' is selected in Q15 \\ \midrule
14 &
  What motivates you to include assertion error messages in your assert methods? &
  Multi-Choice &
  Yes &
  Includes ``Other'' free-text option \\ \midrule
15 &
  Under what conditions would you choose not to include an assertion error message in an assert method? &
  Multi-Choice &
  Yes &
  Includes ``Other'' free-text option \\ \midrule
16 &
  What factors or elements in the source code do you consider when constructing assertion error messages? &
  Multi-Choice &
  Yes &
  Includes ``Other'' free-text option \\ \midrule
17 &
  What are the typical details that you include in an assertion error message? Please rank the following options, with 1 being the most frequently used and 11 being the least frequently used. &
  Ranking &
  Yes &
  Includes ``Other'' free-text option \\ \midrule
18 &
  How do you typically compose an assertion error message? Please rank the following options based on their frequency of usage, with 1 being the most frequently used and 3 being the least frequently used. &
  Ranking &
  Yes &
   \\ \midrule
19 &
  How would you determine if an assertion error message is of poor quality? Please rank the following options based on your evaluation criteria, with 1 being the most indicative of poor quality and 8 being the least indicative of poor quality. &
  Ranking &
  Yes &
  Includes ``Other'' free-text option \\ \midrule
20 &
  What are the common challenges you encounter when writing and maintaining assertion error messages? &
  Multi-Choice &
  Yes &
  Includes ``Other'' free-text option \\ \midrule
21 &
  How often do you review and update assertion error messages to ensure they remain accurate and relevant? Please rank the following options, with 1 being the most frequently used and 6 being the least frequently used. &
  Ranking &
  Yes &
  Includes ``Other'' free-text option \\ \midrule
22 &
  How do you incorporate assertion error messages into the debugging process of test case failures? &
  Multi-Choice &
  Yes &
  Includes ``Other'' free-text option \\ 
  \bottomrule
\end{tabular}%
}
\end{table*}

\subsection{Survey Participants}
To ensure that we recruited qualified participants for our study, we utilized LinkedIn \cite{LinkedIn} -- the leading professional social media network with more than 1 billion users across various industries and fields, including information and technology \cite{LinkedInAboutUs,aguado2019linkedin}. This type of purposive sampling allowed us to identify and invite individuals with the necessary skills and experience required for our study \cite{tongco2007purposive,Baltes2022}. Prior research studies have used LinkedIn to recruit practitioners, and shown that LinkedIn is an ideal source for identifying a diverse and experienced set of survey candidates \cite{Mello2015,mirabeau2013utility,Kanij2013,Daka2014,Ozkaya2020}.

To find potential participants, we searched for practitioners with unit testing experience. Specifically, we performed a people search looking for practitioners having one or more of the following keywords: ``unit testing,'' ``unit tests,'' ``JUnit,'' ``NUnit,'' ``TestNG,'' ``MSTest,'' ``pytest,'' ``PHPUnit,'' ``Mocha,'' ``XCTest.''	For each keyword, we selected 100 random practitioners to connect and message, which roughly equates to a confidence level of 95\% and 10\% margin of error, for a total of 1.000 unique practitioners\footnote{Due to budget constraints, we could not use LinkedIn's premium version. As a result, we had limitations on the number of searches and connections we could make in a given period.}. 
To ensure that our sample consisted of practitioners with relevant expertise, we manually reviewed each potential participant's profile for experience with unit testing. This involved examining the description associated with their employment history and projects to look for instances indicating unit testing experience. We excluded profiles where the testing keywords only appeared under the Courses section, as our objective is to recruit participants with practical experience.

Using LinkedIn's messaging feature, we sent each selected participant an invitation and a link to the online Qualtrics survey. Before starting the survey, participants were shown an informed consent page that outlined the study's purpose, procedures, risks, and benefits. They had to actively agree to this before proceeding with the survey questions. Participation was completely voluntary, and no compensation or other benefits were offered.

\subsection{Pilot Study}
As best practice, we conducted a pilot study before distributing the questionnaire \cite{linaaker2015guidelines}. We worked with six practitioners to identify potential issues with the questionnaire\footnote{The pilot participants were recruited from the author's professional network. They were instructed to review the survey questionnaire carefully and provide unbiased and constructive feedback.}. From their feedback, we identified a few areas that needed improvement. Specifically, we found a flaw with the branching logic: some questions required rewording to make them easier to understand, and a few questions needed re-ordering. After making these adjustments, we submitted our study for IRB approval, and once it was approved, we made it publicly available. 

\subsection{Data Analysis}
We use quantitative and qualitative techniques to analyze the survey data \cite{Wagner2020}. Our quantitative analysis includes applying statistical methods, while for our qualitative analysis, we manually reviewed the free-text responses provided by participants and identified common themes and categories to gain insights into their thoughts and opinions. To ensure reliability, two authors independently reviewed and categorized the free-text responses, discussed any discrepancies, and reached a consensus. When reporting qualitative results, we include, where applicable, representative quotes to support our findings.

\section{Results}
\label{Section:experiment_results}
This section presents our RQ results\footnote{Data in some tables is limited due to space constraints. However, all data can be found online at \cite{Dataset}.}$^{,}$\footnote{Pilot participants' responses were used to identify potential issues with the questionnaire and are not included in our RQ results.}.
We first report the number of responses received and participant demographics before addressing our RQs.

\subsection*{\textbf{Survey Responses}}
The survey was open to the public from September 2023 to March 2024, and we received 217 responses during this period. However, not all participants provided answers to all questions. To ensure consistency in our analysis of results, we only considered participants who answered all mandatory questions, resulting in 138 valid responses. Hence, our reporting of results is limited to these 138 survey responses.

\subsection*{\textbf{Participant Demographics}}
We divide our reporting of participant demographics into two parts; first, we report on general demographic details, followed by work experience.

\subsubsection*{General Demographics}
Our survey received responses from participants residing in 30 countries across Africa, Asia, Oceania, Europe, and North and South America. The top three countries where our participants reside are the United States (55 respondents or 40\% of the total), Brazil (10 respondents or 7\%), and Canada (8 respondents or 6\%). Among the respondents, 114 (83\%) identify as male, while 20 (14\%) identify as female, and 4 (3\%) identify as non-binary/third gender or prefer not to disclose. When it comes to age, most participants fall within the age range 25-34 years (51 respondents or 37\%) and 35-44 years (50 or 36\%). Finally, in terms of their primary language preference for both spoken and written communication, 97 (70\%) respondents selected English, followed by Portuguese with 10 respondents (7\%), while the remaining 22\% specified other languages.

\subsubsection*{Work Experience}
The majority of the respondents (58 or 42\%) report more than 10 years of general programming experience, while 43 (31\%) respondents report between 6 to 10 years, and 30 (22\%) report between 3 to 5 years of programming experience. In terms of employment status, the top three responses received are: working full-time (112 respondents or 81\%), unemployed and looking for work (12 or 9\%), and working part-time (6 or 4\%). When asked to describe their role, the top two responses are Software Engineer/Developer (79 respondents or 57\%) and QA/Test Engineer (31 or 22\%). With regards to unit testing, as shown in Table \ref{Table:demographic_UnitTestingFamiliarity}, all participants are aware of unit testing, with only 5 (4\%) respondents reporting ``Slightly familiar,'' which is on the lower end of the Likert scale. Likewise, in Table \ref{Table:demographic_UnitTestingJob}, we observe that most participants report that writing unit tests is part of their job duties. Finally, we also observe a range of unit testing frameworks the participants are familiar with, the top five being JUnit (74	or 54\%), NUnit (14 or 10\%), PyTest (14 or 10\%), Jest (9 or 7\%), and TestNG (8 or 6\%).

\begin{table}
\centering
\caption{Respondents' familiarity with unit testing.}
\label{Table:demographic_UnitTestingFamiliarity}
\normalsize
\begin{tabular}{@{}lrr@{}}
\toprule
\multicolumn{1}{c}{\textbf{Answer Options}} & \multicolumn{1}{c}{\textbf{Count}} & \multicolumn{1}{c}{\textbf{Percentage}} \\ \midrule
Extremely familiar  & 60 & 43\% \\
Very familiar       & 49 & 36\% \\
Moderately familiar & 24 & 17\% \\
Slightly familiar   & 5  & 4\%  \\
Not familiar at all & 0  & 0\%  \\ \bottomrule
\end{tabular}%
\end{table}

\begin{table}
\centering
\caption{Extent to which writing unit tests part of job duties}
\label{Table:demographic_UnitTestingJob}
\normalsize
\begin{tabular}{@{}lrr@{}}
\toprule
\multicolumn{1}{c}{\textbf{Answer Options}} & \multicolumn{1}{c}{\textbf{Count}} & \multicolumn{1}{c}{\textbf{Percentage}} \\ \midrule
All of the time  & 26 & 19\% \\
Most of the time & 23 & 17\% \\
A lot            & 30 & 22\% \\
Somewhat         & 32 & 23\% \\
Very little      & 18 & 13\% \\
Not at all       & 6  & 4\%  \\
Not Applicable   & 3  & 2\%  \\ \bottomrule
\end{tabular}%
\end{table}

The above demographic data demonstrates that although our sample size consists of 138 survey respondents, the participants are diverse and representative, ensuring that the findings apply to a broad range of individuals within the software engineering community.

\subsection*{\RQA}
This RQ comprises two sub-RQs that aim to understand the current frequency motivations behind including or excluding assertion messages in test methods.

\subsubsection*{\RQAa}
\begin{figure}
    \centering
    \includegraphics[width=1\linewidth]{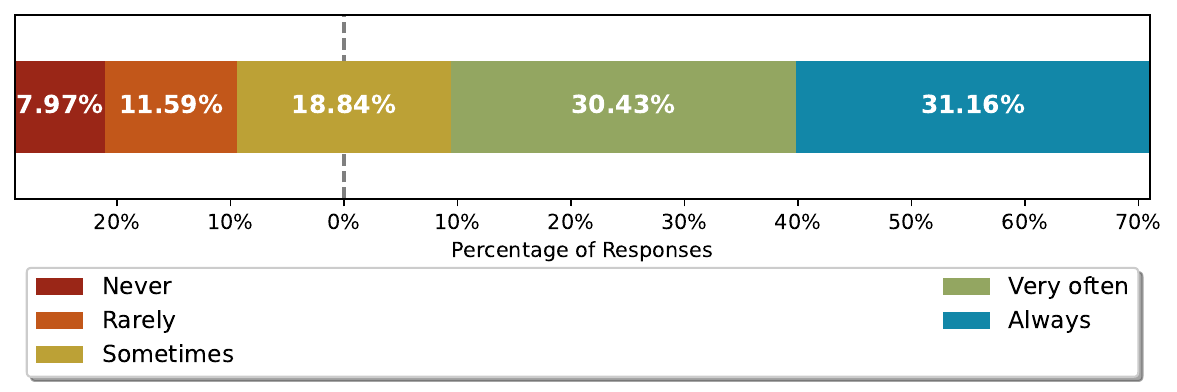}
    \caption{Frequency of including messages in assertion methods.}
    \label{Figure:rq1a_likert}
\end{figure}

To answer this RQ, we examine the responses to survey questions \#12 and \#13. In \#12, respondents report how frequently they include assertion messages. As shown in Figure \ref{Figure:rq1a_likert}, most respondents report using assertion messages, with 43 (31.16\%) selecting ``Always'' and 42 (30.43\%) selecting ``Very often.'' In contrast, 16 (11.59\%) report ``Rarely,'' and 11 (7.97\%) report ``Never.''

In survey question \#13, we ask respondents who selected ``Rarely'' and ``Never'' to explain their answer choices. Two of the authors reviewed the 27 free-text responses to this survey question to identify common themes. Our analysis yields five high-level categories, which we describe below:

\begin{itemize}
    \item \textbf{Redundancy} - Most respondents state that the test case failure itself implies the reason. Major sub-categories include:
    \begin{itemize}
        \item \textbf{Intuitive Assertions} - The assertion methods themselves are often self-explanatory -- e.g., ``\textit{If I am asserting equals and the test fails, I know the 2 values are not equal.}''
        \item \textbf{Satisfactory Default Message} - The default assertion message provided by the testing framework is sufficient to identify the cause of the failure -- e.g., ``\textit{Default error messages are clear enough whenever a test case fails.}''
    \end{itemize}
    \item \textbf{Clean Code} - Writing tests that are simple, understandable, and maintainable helps in understanding the failure of the test -- e.g., ``\textit{Typically, I keep the test logic clean enough that figuring out why a test failed is pretty easy.}''
    \item \textbf{Test Method Name} - The context of the test's failure can be easily understood from a descriptive test method name -- e.g., ``\textit{If you name your test method descriptively, the cause of the failed assertion will be evident.}''
    \item \textbf{Unfamiliarity and Practices} - One respondent mentioned not being familiar with assertion messages -- e.g., ``\textit{Not familiar with how to implement.}'' Additionally, two mentioned it was not standard practice in their workplace --  e.g., ``\textit{In my entire commercial career never used/seen those.}''
    \item \textbf{Framework Limitation} - One respondent mentioned that including messages in assertion methods is not easy with their testing framework (e.g., Jest - https://jestjs.io/).
\end{itemize}

\subsubsection*{\RQAb}
In this RQ, we examine the responses to survey questions \#14 and \#15. In \#14, we are interested in understanding the reasons for using assertion messages, while \#15, on the other hand, focuses on the concerns with using these messages. Both questions allow respondents to select multiple options from a predefined list, with an additional choice to enter their own free text labeled under `Other'. 

We first examine the reasons for including an assertion message. Since this is a multiple-choice question, we will examine the frequency of each answer option and then focus on the most common combinations of answers. From Table \ref{Table:rq1b_includeInduvidual}, we observe that the frequent need for having messages is to help troubleshoot test case failures, followed by using them to document the expected behavior of the test case. Examining the `Other' free text responses, we observe that messages are utilized to display variable values (e.g., ``\textit{Preserving values after execution. NUnit ex: \$`Expected 3 but was {result.Value}'}''). In addition, some respondents mentioned that they only typically use messages if the test case is complex.

Our analysis reveals that, on average, respondents select 2.34 answer options for this multiple-choice question. The most common two-answer option is `Help troubleshoot test case failures' and `Improve understandability of the test method,' occurring 73 times. The most common three-answer option is `Help troubleshoot test case failures,' `Improve understandability of the test method,' and `To serve as documentation for the test case,' occurring 41 times.

Moving on, we examine the reasons for not using assertion messages and follow the same approach described above. Table \ref{Table:rq1b_excludeInduvidual} indicates that respondents tend to prefer the default assertion messages when a test case fails. Additionally, by not including a message, their code becomes simple, clear, and concise, which can help reduce development time.
When analyzing the `Other' text entries, we encounter more qualitative data than the prior survey question. We observe some themes, similar to RQ1.1; specifically, the use of high-quality test method names (e.g., ``\textit{When the method name fully explains the assertion}''), intuitive assertions (e.g., ``\textit{Don't state the obvious}''), not a project requirement (e.g., ``\textit{customer does not want}''), reliance on default framework messages/functionality (e.g., ``\textit{...stack trace will be included and enough to debug it}''), and framework incompatibility. Interestingly, eleven respondents used this `Other' option to indicate that they always use assertion messages (e.g., ``\textit{I always add them}'').

With the average number of answer choices selected by respondents for this question being 1.63, we next look at the common answer combinations. `I find the information present in the test execution log sufficient' and 'I prefer to keep my assert methods simple, clear and concise' is the most common two-answer combination, occurring 29 times. The most common three-answer combination is `I find the information present in the test execution log sufficient,' `I prefer to keep my assert methods simple, clear and concise,' and `Reduce development time,' occurring 7 times.

\begin{table}
\centering
\caption{Frequency of occurrence for each of the answer options for question \#14}
\label{Table:rq1b_includeInduvidual}
\resizebox{\columnwidth}{!}{%
\begin{tabular}{@{}p{0.4\linewidth}rr@{}} 
\toprule
\multicolumn{1}{c}{\textbf{Individual Answer Option}} & \multicolumn{1}{c}{\textbf{Count}} & \multicolumn{1}{c}{\textbf{Percentage}} \\ \midrule
Help troubleshoot test case failures         & 105 & 32.51\%  \\ \midrule
Improve understandability of the test method & 86  & 26.63\% \\ \midrule
To serve as documentation for the test case  & 61  & 18.89\%  \\ \midrule
Adherence to coding standards                & 39  & 12.07\%  \\ \midrule
Feedback from code reviews                   & 21  & 6.50\%  \\ \midrule
Other                                        & 11  & 3.41\%  \\ \midrule
\multicolumn{1}{r}{\textit{\textbf{Total}}}           & \textit{\textbf{323}}              & \textit{\textbf{100.00\%}}              \\ \bottomrule
\end{tabular}%
}
\end{table}

\begin{table}
\centering
\caption{Frequency of occurrence for each of the answer options for question \#15}
\label{Table:rq1b_excludeInduvidual}
\resizebox{\columnwidth}{!}{%
\begin{tabular}{@{}p{0.4\linewidth}rr@{}} 
\toprule
\multicolumn{1}{c}{\textbf{Individual Answer Option}} & \multicolumn{1}{c}{\textbf{Count}} & \multicolumn{1}{c}{\textbf{Percentage}} \\ \midrule
I find the information present in the test execution log sufficient & 55 & 24.44\%  \\ \midrule
I prefer to keep my assert methods simple, clear and concise        & 52 & 23.11\% \\ \midrule
Reduce development time                                             & 37 & 16.44\% \\ \midrule
Other                                                               & 35 & 15.56\% \\ \midrule
Reduce additional rework when modifying test case                   & 23 & 10.22\% \\ \midrule
I am unaware that an error message is an option                     & 12 & 5.33\%  \\ \midrule
I prefer to use code comments instead of assertion error messages   & 11 & 4.89\%  \\ \midrule
\multicolumn{1}{r}{\textit{\textbf{Total}}}           & \textit{\textbf{225}}              & \textit{\textbf{100.00\%}}              \\ \bottomrule
\end{tabular}%
}
\end{table}

\begin{tcolorbox}[top=0.5pt,bottom=0.5pt,left=1pt,right=1pt]
\textbf{Summary for RQ1.}
While most respondents find assertion messages useful, not all of them use messages in every assertion method. These messages are often used to troubleshoot failures, improve understandability, and serve as documentation. Reasons for not using them include redundancy due to intuitive assertion statements, adhering to clean code principles and descriptive test names, and preferences for default framework messages.
\end{tcolorbox}

\subsection*{\RQB}
We approach this RQ from multiple angles. First, from survey question \#16, we examine the elements in the code that influence the structure and wording of the message. Then, from question \#17, we examine the typical details included in the message, and finally, from \#18, we examine message composition style.

\subsubsection*{Source Code Elements}
Survey question \#16 is multi-choice, with an `Other' free text option. It aims to identify the specific code elements and contextual factors that practitioners consider when creating assertion messages. Table \ref{Table:rq2_elements} shows the frequency occurrence of each answer option. From this table, we observe that the assertion method plays an important role in formulating the message, with 69 (or 20\%) respondents relying on the identifiers used in expected and actual arguments and 67 (or 19.42\%) respondents relying on the type of assert call. The least influence on the message comes from other assertions in the same test case.

As this is a multi-choice question, respondents selected, on average, 2.5 answer options. The most commonly chosen combination of two answer options is `The type of assert method containing the assertion error message' and `The variable or functions that are utilized as the actual/expected arguments in the assert method,' which occurs 37 times, while the next highest combination in `Exception/Error handling code within the test method' and `The type of assert method containing the assertion error message,' which occurs 30 times. Lastly, the most common three-option combination is `The name of the test class,' `The name of the test method,' and `The type of assert method containing the assertion error message,' occurring 18 times.

Lastly, examining the free text of the `Other' option reveals varying practices. Some practitioners consider factors such as the intent of the functionality being tested and external dependencies. Others regard assertion messages as unnecessary, considering tests should be self-documenting or that details should be added only when the failure is not apparent.

\subsubsection*{Details Included in a Message}
This survey question (\#17) asks respondents to rank eleven answer options related to various details that can be included in a message. To determine the overall ranking, we use the Borda Count method \cite{Saari2023}, which is specifically designed for ranked preference data. This method assigns points to each option based on its rank in each respondent's list. The higher the rank, the more points an option receives. Table \ref{Table:rq2_borda} shows the top five preferred types of information to be included in the assertion message. The Borda Score column shows the total points each option received based on the respondents' rankings. The "Prop." (i.e., proportion) column compares each option's score to the maximum possible score, which is 1,518	(138 respondents × 11 points for a first-place rank).

Findings show that including a failure description is the top-ranked option, followed by the expected outcome and actual outcome in positions two and three, respectively. Moreover, the top-ranked item receives about 92\% of the maximum possible points (i.e., proportion), indicating a strong consensus among most respondents about preferring a descriptive failure message compared to the other provided answer options. Further, though not shown in the table, the last-ranked item is `Other,' with a score of 206 and a proportion value of 0.14.

\subsubsection*{Message Style}
Prior work \cite{Takebayashi2023NLBSE} on open-source projects identified three styles of message compositions. These styles are: messages composed solely of string literals (e.g., \texttt{assertEquals(``Values not equal'',a,b)}), messages consisting only of a variable/method call (e.g., \texttt{assertEquals(r.getValue(), a, b)}), and messages that are a combination of both (e.g., \texttt{assertFalse(``too big: '' + a, a>5)}).

As part of answering this RQ, we asked respondents to rank their preferred style (question \#18). Like the prior survey question, we use the Borda Count method to analyze the responses, and the results are shown in Table \ref{Table:rq2_style}. Similar to \cite{Takebayashi2023NLBSE}, our results show participants preferring messages composed of a single string literal over the other two styles. However, in contrast to the prior study, survey respondents preferred identifier-only messages over combination messages.

\begin{table}
\centering
\caption{Frequency of occurrence for each of the answer options for question \#16}
\label{Table:rq2_elements}
\resizebox{\columnwidth}{!}{%
\begin{tabular}{@{}p{0.4\linewidth}rr@{}} 
\toprule
\multicolumn{1}{c}{\textbf{Individual Answer Option}} &
  \multicolumn{1}{c}{\textbf{Count}} &
  \multicolumn{1}{c}{\textbf{Percentage}} \\ \midrule
The variable or functions that are utilized as the actual/expected arguments in the assert method &
  69 &
  20.00\% \\ \midrule
The type of assert method containing the assertion error message (e.g., assertTrue, assertEquals, fail, etc.) &
  67 &
  19.42\% \\ \midrule
Exception/Error handling code within the test method                              & 56                    & 16.23\%                    \\ \midrule
The name of the test method                                                       & 49                    & 14.20\%                    \\ \midrule
The source/production method under test                                           & 42                    & 12.17\%                    \\ \midrule
The name of the test class                                                        & 31                    & 8.99\%                     \\ \midrule
Other                                                                             & 20                    & 5.80\%                     \\ \midrule
Other assert methods, if any, that are also contained within the same test method & 11                    & 3.19\%                     \\ \midrule
\multicolumn{1}{r}{\textit{\textbf{Total}}}                                       & \textit{\textbf{345}} & \textit{\textbf{100.00\%}} \\ \bottomrule
\end{tabular}%
}
\end{table}

\begin{table}
\centering
\caption{Top five details included in an assertion error message ranked by the Borda count method.}
\label{Table:rq2_borda}
\normalsize
\begin{tabular}{@{}rlrr@{}}
\toprule
\multicolumn{1}{c}{\textbf{Rank}}&\multicolumn{1}{c}{\textbf{Answer Option}} & \multicolumn{1}{c}{\textbf{Borda Score}} & \multicolumn{1}{c}{\textbf{Prop.}} \\ \midrule
1 & Description of the failure & 1,397 & 0.92 \\ \midrule
2 & Expected outcome           & 1,244 & 0.82 \\ \midrule
3 & Actual outcome             & 1,137 & 0.75 \\ \midrule
4 & Test method name           & 933  & 0.61 \\ \midrule
5 & Suggestions for resolution & 904  & 0.60 \\ \bottomrule
\end{tabular}%
\end{table}

\begin{table}
\centering
\caption{Message style composition ranked by the Borda count method.}
\label{Table:rq2_style}
\resizebox{\columnwidth}{!}{%
\normalsize
\begin{tabular}{@{}rp{0.48\linewidth}rr@{}} 
\toprule
\multicolumn{1}{c}{\textbf{Rank}} &
\multicolumn{1}{c}{\textbf{Answer Option}}                                  & \multicolumn{1}{c}{\textbf{Borda Score}} & \multicolumn{1}{c}{\textbf{Prop.}} \\ \hline
1 & Composed only of a String literal              & 359 & 0.87 \\ \midrule
2 & Composed of one or more variables/method calls & 250 & 0.60 \\ \midrule
3 & Combination of String literals and variables/method calls & 219                                      & 0.53 \\ \bottomrule     
\end{tabular}%
}
\end{table}

\begin{tcolorbox}[top=0.5pt,bottom=0.5pt,left=1pt,right=1pt]
\textbf{Summary for RQ2.}
Assertion messages are usually constructed by considering the type of assertion method, the identifiers used in the expected and actual arguments, and the names of the test class and method. The message usually contains information such as the failure description and the expected and actual values, and is often constructed using a single string literal. Finally, some practitioners consider that tests should be self-explanatory and that additional information should only be included when the reason for failure is not obvious.
\end{tcolorbox}

\subsection*{\RQC}
For this research question, we use multiple survey questions to gather information. Firstly, we explore the criteria used to identify if a message is of low quality through survey question \#19. Secondly, based on survey question \#20, we identify common difficulties in maintaining assertion messages. Lastly, survey question \#21 helps us understand at what point in time a message is usually reviewed for accuracy.

Starting with question \#19, we use the Borda Count technique to examine how respondents rank factors indicative of poor-quality assertion messages. As per Table \ref{Table:rq3_quality}, the primary indicator of poor-quality messages is the absence of clear information regarding the nature of the failure. This concern received a rating equivalent to 0.93 of the highest possible Borda score, emphasizing its importance in the evaluation process. Following that, assertion messages that fail to provide diagnostic information are considered the second most significant indicator of poor quality, with excessively technical messages coming in third. In last place is the `Other' option with a proportion value of 0.22, with some respondents using this option to mention that the existence of assertion messages can signify that the assertion statement is not self-describing or if ``\textit{the message just repeats the result of the assertion, such as `a and b are not equal'}.''

Moving on, question \#20 asked the participants about the challenges they face in creating and maintaining assertion messages. According to Table \ref{Table:rq3_challenges}, 71 respondents identified that the biggest challenge is crafting messages that accurately describe the nature of the failure. Following closely, 69 respondents highlighted the challenge of balancing message conciseness and descriptiveness. As this is a multi-choice question, participants typically selected around 2.57 options, indicating a diverse range of concerns, with the combination of these two primary challenges frequently occurring together, happening 37 times. The `Other' responses include a lack of documentation and respondents who do not face challenges.

Finally, from survey question \#21, we ask respondents how they prioritize reviewing and updating assertion messages to ensure they remain accurate and relevant. Table \ref{Table:rq3_review} shows the top five priorities based on the Borda Score approach. Updating the message when the test case fails emerges as the clear favorite among the respondents, emphasizing the importance of real-time troubleshooting and maintenance. The next two items—updating messages during code reviews and after updates to production code—are closely contested, indicating a relatively balanced preference between them. Significant changes or enhancements to the tested functionality rank fourth, highlighting the need to keep assertion messages aligned with evolving code requirements. Examining the "Other" responses reveals additional insights, such as updating messages during periods of lower activity and adopting a minimalist approach by only updating assertions when essential.

\begin{table}
\centering
\caption{Top five options for determining the quality of assertion error messages ranked by the Borda count method.}
\label{Table:rq3_quality}
\resizebox{\columnwidth}{!}{%
\normalsize
\begin{tabular}{@{}rp{0.5\linewidth}rr@{}} 
\toprule
\multicolumn{1}{c}{\textbf{Rank}} & \multicolumn{1}{c}{\textbf{Answer Option}} & \multicolumn{1}{c}{\textbf{Borda Score}} & \multicolumn{1}{c}{\textbf{Prop.}} \\ \midrule
1 & The message lacks clarity on why the test failed                         & 1031 & 0.93 \\ \midrule
2 & The message does not include diagnostic data to troubleshoot the failure & 849  & 0.77 \\ \midrule
3 & The message contains excessive technical language/jargon                 & 737  & 0.67 \\ \midrule
4 & The message is too short                                                 & 596  & 0.54 \\ \midrule
5 & The message contains grammatical/punctuation/spelling mistakes           & 567  & 0.51 \\ \bottomrule
\end{tabular}%
}
\end{table}

\begin{table}
\centering
\caption{Frequency of occurrence for each of the answer options for question \#20}
\label{Table:rq3_challenges}
\resizebox{\columnwidth}{!}{%
\begin{tabular}{@{}p{0.5\linewidth}rr@{}} 
\toprule
\multicolumn{1}{c}{\textbf{Individual Answer Option}} & \multicolumn{1}{c}{\textbf{Count}} & \multicolumn{1}{c}{\textbf{Percentage}} \\ \midrule
Making sure that the error messages accurately reflects failure                                              & 71                    & 20.00\%                    \\ \midrule
Striking the right balance between message conciseness and descriptiveness                                   & 69                    & 19.44\%                    \\ \midrule
Keeping the messages up to date and relevant as the codebase evolves                                         & 60                    & 16.90\%                    \\ \midrule
Maintaining consistency in the structure and composition of the message across the test suite in the project & 48                    & 13.52\%                    \\ \midrule
Understanding the intent and context of assert methods written by other developers (such as in legacy code)  & 48                    & 13.52\%                    \\ \midrule
Incorporating sufficient details to help with debugging/troubleshooting failures                             & 45                    & 12.68\%                    \\ \midrule
Other                                                                                                        & 14                    & 3.94\%                     \\ \midrule
\multicolumn{1}{r}{\textit{\textbf{Total}}}                                                                  & \textit{\textbf{355}} & \textit{\textbf{100.00\%}} \\ \bottomrule
\end{tabular}%
}
\end{table}

\begin{table}
\centering
\caption{Top five options for determining when to update assertion error messages ranked by the Borda count method.}
\label{Table:rq3_review}
\resizebox{\columnwidth}{!}{%
\normalsize
\begin{tabular}{@{}rp{0.5\linewidth}rr@{}} 
\toprule
\multicolumn{1}{c}{\textbf{Rank}} & \multicolumn{1}{c}{\textbf{Answer Option}} & \multicolumn{1}{c}{\textbf{Borda Score}} & \multicolumn{1}{c}{\textbf{Prop.}} \\ \midrule
1 & When encountering test failures or issues related to the assert methods            & 638 & 0.77 \\ \midrule
2 & During regular code review processes with team members                             & 603 & 0.73 \\ \midrule
3 & After every update to the production code                                          & 598 & 0.72 \\ \midrule
4 & Whenever there are significant changes or enhancements to the tested functionality & 571 & 0.69 \\ \midrule
5 & I never update assert methods after they are initially implemented                 & 306 & 0.37 \\ \bottomrule
\end{tabular}%
}
\end{table}

\begin{tcolorbox}[top=0.5pt,bottom=0.5pt,left=1pt,right=1pt]
\textbf{Summary for RQ3.}
The primary indicators of low-quality assertion messages are a lack of clear information about the failure and insufficient diagnostic details. Crafting accurate messages and balancing conciseness and descriptiveness can be challenging. To ensure that these messages are relevant and effective, they are usually reviewed and updated when test cases fail, during code reviews, or after changes to the production code. The results underscore the importance of keeping assertion messages in sync with the evolving codebase throughout the development lifecycle.
\end{tcolorbox}

\subsection*{\RQD}
This RQ examines the responses to the multi-choice question \#22. Table \ref{Table:rq4} shows the distribution of answer choices. The high percentage of respondents using assertion messages as a starting point (32.97\%) suggests that these messages are valuable for understanding the nature of the failure. However, the need to apply additional debugging techniques highlights that these messages alone may not always provide sufficient detail or context to pinpoint the root cause. The second most frequent approach (21.15\%) is leveraging debugging tools and IDE features alongside assertion messages, showing the importance of the advanced features and plugins incorporated in IDEs, such as breakpoints, inspections, and watches. Additionally, 14.34\% of respondents combine assertion messages with custom logging statements, enabling them to capture other relevant information about the state of the application and test case leading up to the assertion failure. Moreover, respondents select, on average, 2.02 answer choices, with the most common combination being the use of debugging tools/IDE features together with assertion messages (49 occurrences). 

\begin{table}
\centering
\caption{Frequency of occurrence for each of the answer options for question \#22}
\label{Table:rq4}
\resizebox{\columnwidth}{!}{%
\begin{tabular}{@{}p{0.5\linewidth}rr@{}} 
\toprule
\multicolumn{1}{c}{\textbf{Individual Answer Option}} &
  \multicolumn{1}{c}{\textbf{Count}} &
  \multicolumn{1}{c}{\textbf{Percentage}} \\ \midrule
I utilize the assertion error message as a starting point and then apply additional debugging techniques to explore further &
  92 &
  32.97\% \\ \midrule
I leverage debugging tools or IDE features alongside the assertion error message     & 59                    & 21.15\%                    \\ \midrule
I combine the assertion error message with custom logging statements                 & 40                    & 14.34\%                    \\ \midrule
I collaborate with team members to discuss and analyze assertion failure message     & 32                    & 11.47\%                    \\ \midrule
I rely only on the assertion error message to identify the root cause of the failure & 32                    & 11.47\%                    \\ \midrule
I do not rely on the assertion error message and prefer other techniques to identify the root cause of the failure &
  17 &
  6.09\% \\ \midrule
Other                                                                                & 7                     & 2.51\%                     \\ \midrule
\multicolumn{1}{r}{\textit{\textbf{Total}}}                                          & \textit{\textbf{279}} & \textit{\textbf{100.00\%}} \\ \bottomrule
\end{tabular}%
}
\end{table}

\begin{tcolorbox}[top=0.5pt,bottom=0.5pt,left=1pt,right=1pt]
\textbf{Summary for RQ4.}
Although assertion messages can be helpful in identifying test failures, their effectiveness can be enhanced by utilizing additional debugging techniques such as debugging tools, IDE features, and custom logging statements. By adopting a comprehensive approach that includes these techniques, practitioners can gain a deeper understanding of the root cause of test case failures.
\end{tcolorbox}

\section{Discussion \& Takeaways}
\label{Section:discussion}

Our findings from RQ1 reveal that most practitioners include assertion messages in their test cases; they recognize their value in troubleshooting, test understandability, and documentation. From RQ2, we observe that practitioners consider various elements of the test code when constructing assertion messages, aiming to create meaningful and relevant messages. They prioritize information about the failure, expected and actual outcomes, and favor concise string literals. RQ3 highlights the importance of maintaining the quality and relevance of assertion messages throughout the development lifecycle, with practitioners often struggling to keep messages up-to-date with the changing codebase. Finally, RQ4 reveals that most practitioners often combine assertion messages with other debugging techniques to troubleshoot test case failures. It is worth noting that unit testing is not limited to software engineers or developers, but also includes QA/test engineers, which is reflected in our demographic data. Hence, this study refers to these individuals as practitioners. 

While assertion messages are valuable, our findings show that their inclusion is not a universal practice; practitioners are selective in their use. Some survey respondents argue that assertion messages are redundant if the assertion statement is intuitive, and only utilize them in complex assertions. While this perspective is understandable, it is important to note that redundancy is subjective and context-dependent. In certain situations, an assertion message may provide valuable additional information, even if some practitioners consider it redundant. For example, novice developers joining an established project may find the messages helpful; experienced project developers, on the other hand, maybe more accustomed to the codebase and therefore find assertion messages less necessary. Furthermore, although some practitioners emphasize the importance of descriptive test method names instead of assertion messages, studies show that unit tests often lack descriptive names \cite{Zhang2016}. Assertion messages offer an additional way to document specific details that may be challenging to express in the test name alone; they offer more context about the test without developers needing to open the source code file and read the comments. Moreover, as the codebase evolves, test code often fails to co-evolve with production code \cite{zaidman2011studying}, which can lead to instances where test names may become outdated, leading to a mismatch between the test's purpose and name. In such cases, well-maintained assertion messages become a more reliable form of documentation as they are directly linked to the specific conditions being tested, making it easier to update them when needed.  Assertion messages also help improve test code quality by addressing the Assertion Roulette test smell, which occurs when a test method contains multiple assertions without explanatory messages, making it difficult to understand which assertion caused a failure.

\subsection*{\textbf{Takeaways}} 
\noindent\textbf{Improvements to automated test code generation techniques.} As described in Section \ref{Section:related}, techniques that generate assertion statements tend to ignore the assertion message. Since this code is generated automatically rather than written by developers, and typically lacks comments, the absence of meaningful messages can hinder the understandability and maintainability of the generated test code. Practitioners may struggle to comprehend the intent and expected behavior of the tests, especially when failures occur. Furthermore, recent work by Siddiq et al. \cite{Siddiq2024} shows that code-generation tools based on large language models (LLMs) struggle to generate compilable test code and generally achieve lower coverage, especially for real-world code. They also found that LLM-generated test code is more likely to suffer from test smells like Assertion Roulette. Researchers and tool vendors should enhance automated test generation techniques to incorporate the creation of high-quality assertion messages that support troubleshooting and serve as documentation. The findings and insights from our research can provide guidance and a foundation for improving assertion message generation in automated test code creation tools.

\vspace{0.5mm}
\noindent\textbf{Need for tooling to appraise and refactor assertion messages.} Similar to the previous point, there is a need for IDE and tool vendors to develop techniques to evaluate and provide real-time feedback on the quality of existing messages based on factors such as clarity, conciseness, relevance to the associated test case, and consistency across the test suite. For example, our RQ1 findings reveal the reasons for omitting messages; tooling can help highlight instances where these reasons may not be justified and suggest adding meaningful messages. Furthermore, based on our RQ3 findings, these tools could detect modifications to a test's assertions and prompt practitioners to review and update the corresponding messages. This would ensure that the messages remain accurate and helpful throughout the lifecycle of the test suite, similar to work that detects obsolete comments \cite{liu2021just}.

\vspace{0.5mm}
\noindent\textbf{Integrating assertion messages into the debugging workflow.} Findings from RQ4 show that assertion messages are useful for understanding test failures, especially when combined with other debugging techniques. This provides an opportunity for IDE vendors to explore ways to seamlessly connect assertion messages with built-in debugging features, offering practitioners a more efficient troubleshooting experience. IDEs and testing frameworks can provide suggestions based on the content of the message, such as identifying potential causes of the failure or recommending possible solutions and relevant documentation. For instance, Zhang et al. \cite{Zhang2011} revealed that developers spent 14\% less time understanding bugs when failed tests included inferred explanatory code comments. 

\vspace{0.5mm}
\noindent\textbf{Establishment of guidelines and best practices.} To promote the effective use of assertion messages, the research community should create a comprehensive catalog of best practices based on mining software repositories and conducting surveys with experienced practitioners. The catalog can range from techniques for crafting high-quality messages, as discussed in RQ2 and RQ3, to troubleshooting strategies, as reported in RQ4. For instance, Spadini et al. \cite{Spadini2018Review} report that assertions are elements that practitioners consider when conducting code reviews, but they do not specifically mention if or how they evaluate assertion messages. Therefore, an inclusive catalog can provide valuable insights and guidelines for practitioners to create impactful assertion messages. Additionally, educators should incorporate these best practices into their curricula to ensure students are equipped with the skills and mindset to write maintenance-friendly test cases. 

\vspace{0.5mm}
\noindent\textbf{Advancing research into identifier naming in test code.} Although this study and survey do not specifically examine identifier naming, the respondents emphasize the significance of having high-quality names, particularly for test method names, to convey the test's purpose. This is evident in the responses to RQ1 and RQ2. Interestingly, practitioners believe that both identifiers and assertion messages solve some of the same problems. A good identifier name can reduce the need for an assertion message. They also think that a good identifier name can help them create a strong assertion message. A closer look at these phenomena could help determine when an assertion message is necessary and what information to include so that it complements the information provided by the method name or other surrounding identifiers. Finally, the importance of identifier naming in test code is not unique to the survey respondents in this study. Alsuhaibani et al. \cite{Alsuhaibani2021} also highlights the need for naming standards for test cases based on responses to their survey on method naming. While there has been prior work on test method naming \cite{Peruma2021,Wu2020,Lin2019,Daka2017,Zhang2016}, the relationship between assertion messages and identifier names is one area that is lacking research.

\section{Threats To Validity}
\label{Section:threats}
Although we limited participant selection for our survey to LinkedIn, we manually reviewed the profiles of every invited participant to ensure that they had actual experience using a unit testing framework in a software project, rather than just listing it as a skill they acquired from a course. While practical, this technique has some limitations. Specifically, as the profile is user-generated, we cannot verify the information, and there might be other suitable candidates we did not invite due to missing/incomplete profile information. However, this threat associated with user-reported information is common in all survey-based studies. Despite these limitations, our participants were diverse and representative, as evidenced by their demographics in Section \ref{Section:experiment_results}. When identifying potential participants to invite for our study, it is possible that some level of selection bias may exist and could affect the generalizability of the findings. Additionally, although only 138 participants completed all required survey questions, they reported familiarity with unit testing and experience with testing frameworks. 

Some survey questions consist of single or multi-choice responses, which might seem restrictive. Therefore, to ensure that their answers are not influenced or restricted, we provided an `Other' answer option and an opportunity for them to provide additional details in a free-text response. Additionally, some questions may appear to be biased towards the usefulness of assertion messages. To address this, we have included questions about reasons for not using assertion messages and the challenges faced, along with an `Other' free-text option.
Finally, the anonymity of our survey allowed participants to provide honest and unbiased responses.

\section{Conclusion \& Future Work}
\label{Section:conclusion}
Assertion statements are vital to unit testing. They help practitioners verify the expected behavior of the system under test and allow custom messages to aid in troubleshooting failing tests.
In this study, we survey 138 professional software practitioners about their perception and usage of assertion messages in unit tests. Our findings show that most respondents recognize the value of assertion messages in troubleshooting failures, improving test understandability, and serving as documentation. To ensure that these messages are effective, they should include information about the failure, provide sufficient diagnostic details, and be in sync with the evolving codebase. 
However, some practitioners choose to omit these messages due to redundancy, reliance on default framework messages, or a preference for descriptive test names.
Future work in this area includes conducting a large-scale empirical study that examines message quality to identify techniques for automatically appraising and generating meaningful messages.

\bibliographystyle{ieeetr}
\bibliography{main}
\end{document}